# Iterative reconstruction of SiPM light response functions in a square-shaped compact gamma camera

**A. Morozov,**[a,b,*] **, F. Alves,**[c] **J. Marcos,**[a,b] **R. Martins,**[a,b] **L. Pereira,**[a,b] **V. Solovov**[a] **and V. Chepel**[a,b]

[a] *LIP-Coimbra,*
*Department of Physics, University of Coimbra, Coimbra, Portugal*

[b] *Department of Physics,*
*University of Coimbra, Coimbra, Portugal*

[c] *ICNAS,*
*University of Coimbra, Coimbra, Portugal*

* *E-mail:* `andrei@coimbra.lip.pt`

## Abstract

Compact gamma cameras with a square-shaped monolithic scintillator crystal and an array of silicon photomultipliers (SiPMs) are actively being developed for applications in areas such as small animal imaging, cancer diagnostics and radiotracer guided surgery. Statistical methods of position reconstruction, which are potentially superior to the traditional centroid method, require accurate knowledge of the spatial response of each photomultiplier. Using both Monte Carlo simulations and experimental data obtained with a camera prototype, we show that the spatial response of all photomultipliers (light response functions) can be parameterized with axially symmetric functions obtained iteratively from flood field irradiation data. The study was performed with a camera prototype equipped with a 30 x 30 x 2 mm$^3$ LYSO crystal and an 8 x 8 array of SiPMs for 140 keV gamma rays. The simulations demonstrate that the images, reconstructed with the maximum likelihood method using the response obtained with the iterative approach, exhibit only minor distortions: the average difference between the reconstructed and the true positions in X and Y directions does not exceed 0.2 mm in the central area of 22 x 22 mm$^2$ and 0.4 mm at the periphery of the camera. A similar level of image distortions is shown experimentally with the camera prototype.

# Contents

# 1 Introduction

During the last decades, a growing interest in molecular imaging techniques was observed in many areas including preclinical studies (e.g. [1, 2]), early cancer diagnostics (e.g. [3, 4]) and surgery (e.g. [3, 5]). In these areas, imaging often requires gamma cameras with high spatial resolution and sensitivity, while limiting at the same time the maximum size of the detector (for example, systems for small animal imaging or hand-held and intra-body probes).

The demand for such specialized instrumentation resulted in active development of a new generation of compact imaging systems with high spatial resolution and sensitivity (e.g. [2, 6-10]). Until recent years, compact gamma cameras were usually equipped with position-sensitive photomultiplier tubes (PMTs), but after the introduction of silicon photomultipliers (SiPMs), the use of this type of photosensors becomes increasingly common. SiPMs, while offering the same (or even higher) photon detection efficiency compared to PMTs, have additional advantages such as smaller thickness-to-sensitive-area ratio, low operating voltage and insensitivity to magnetic field (e.g. [11]).

Similarly to clinical gamma cameras, the design of a compact gamma camera usually includes a scintillator coupled through a flat lightguide to an array of photomultipliers. The scintillator can be either an array of crystals, optically insulated from each other (e.g. [6]) or a monolithic scintillator (e.g. [8]). The advantage of the former is a relative simplicity of the spatial calibration procedures since the scintillation from a single event is well localized (confined to a single crystal of the array) and results in a well-defined pattern of the photosensor signals. However, the second approach, besides being technologically simpler, can provide better spatial resolution which, in this case, is not limited by the pitch of the crystal array.

Event position reconstruction is usually performed with the traditional centroid method, or, alternatively, using statistical reconstruction methods [12, 13]. For each event, statistical methods search for the position (and, optionally, the event energy) which results in the best match between the measured photosensor signals and those provided by a mathematical model of the detector. Therefore, these methods require knowledge of the response of each individual photosensor as a function of the event position (so-called light response function, LRF). The statistical methods can, in principle, give more accurate position reconstruction compared to the centroid and offer significantly better capability to discriminate noise and multiple events [13].

The LRFs can be obtained from Monte Carlo simulations with the accuracy depending very much on the assumptions of the simulation model and the knowledge of the relevant physical properties. Alternatively, the LRFs can be calculated from the calibration data acquired in a scan of the detector field of view with a pencil beam [10]. However, such calibrations are time consuming (see, for example, [14]) and are difficult to perform *in situ* as they normally require direct access to the crystal. Therefore, it is impractical to perform such calibrations on a regular basis or apply them for devices employing many small cameras as, for example, SPECT scanners for small animals.

A new technique to obtain LRFs has been recently introduced [15] and applied later for clinical gamma cameras [16]: the LRFs were obtained from flood field irradiation data using an iterative procedure. It was also shown that high uniformity in the distribution of the events over the detector's field of view is not required [16], thus, potentially, opening a possibility to use background radioactivity events to perform detector calibration.

In all our previous studies where the iterative LRF reconstruction was successfully applied [15-17], the spatial response of the photomultipliers was axially symmetric: the LRFs could be safely considered to be functions of only the distance between the sensor's center and the source position, projected on the sensor plane. This approach seems to be ill-suited for compact gamma cameras with square-shaped photosensors and a square-shaped monolithic scintillator. However, in this study we show that a relatively low level of distortions in the reconstructed images can be achieved using maximum likelihood reconstruction with axially symmetric LRFs. Then we demonstrate, using both simulations and experimental data recorded with a camera prototype, that the iterative LRF reconstruction can be successfully applied for this type of gamma cameras.

## 2 Methods

### 2.1 Iterative method of LRF reconstruction

The iterative LRF reconstruction method has already been described in detail in our previous publications [15-17], therefore, only a brief description is given here. The method requires two datasets for the same set of events, distributed over the field of view of the detector: one with the signals of the photosensors and the other one with the estimates of the event positions. The iterative cycle consists of two stages: during the first stage the signals and the event positions are used to evaluate the LRFs of the sensors (LRF reconstruction stage). In the second stage, the new estimated event positions are obtained with a statistical reconstruction method using these LRFs (position reconstruction stage). The cycle is repeated until convergence is reached: one can directly monitor the variation of the LRF profiles from iteration to iteration, or observe a parameter describing how well the reconstructed LRFs represent the provided sensor signals. For example, a parameter proportional to the chi-square of event position reconstruction averaged over all events was used in [16].

If it is possible to provide an initial guess on the LRFs, the cycle can start from the position reconstruction stage. For instance, when processing experimental data, the LRFs can be obtained from the detector simulations [16]. Alternatively, the cycle can start from the LRF reconstruction stage, using, for example, position estimates given by the centroid reconstruction [15].

Several techniques have been developed (see, e.g., [16]) to improve the convergence speed and help to avoid convergence to a local minimum. One of them is to apply a random shift to the reconstructed positions after they were calculated in an iteration, for example, by adding random values, sampled from a Gaussian distribution with zero mean and a small sigma (compared to the inter-sensor distance) to x and y coordinates of each event. In the rest of the article we refer to this technique as "blurring". The blurring is especially important during the first iterations, when due to inaccuracy in the LRF profiles the reconstructed events positions can form artifacts in the reconstructed images, persistent from one iteration to another.

Another technique, usually applied during the first iterations, is to use an LRF parameterization scheme in which all sensors share the same LRF profile, but have individual scaling factors to account for their relative gains. This technique allows to establish, during these first iterations, the general profile of the LRFs which is characteristic to the sensor geometry. In further iterations, the parameterization scheme can be switched to one in which each sensor has individual LRF in order to take into account the differences between the individual sensors (e.g., defined by the position of the sensor in the detector).

It is also important, before each LRF reconstruction stage, to filter out events with unrealistic reconstructed position (and/or energy) or those with a value of a certain goodness-of-fit parameter (e.g., chi-square of the reconstruction) beyond a given threshold. If these events are not removed, they may result in distortions in the reconstructed LRF profiles, which, in turn, can produce persistent artifacts in the images.

## 2.2 Experimental prototype

The camera prototype has a 2 mm thick, 30 x 30 mm$^2$ LYSO(Ce) scintillator from Epic-Crystal [18]. On one side, the crystal faces a 1 mm thick polytetrafluoroethylene (PTFE) reflector. On the other side, it is coupled to a 1.5 mm thick, 31 x 32.5 mm$^2$ acrylic glass lightguide. The side walls of the crystal and the lightguide are painted with a black acrylic paint (refractive index of 1.5).

The camera is equipped with four ArraySB-4 silicon photomultiplier arrays from SensL. An ArraySB-4 unit holds 4 x 4 MicroSB-30035 SiPMs sensors [19], packed with a 3.17 mm pitch inside a ceramic holder. In this study, the SiPMs were operated at 2.5 V above the breakdown voltage. The manufacturer provides a value of photon detection efficiency of 31% at this voltage with the sensitivity peak at 420 nm. The SiPMs, protected with an epoxy layer, are located 0.6 mm below the rim of the ceramic holder, which is put in contact with the lightguide. The array elements are installed on a printed circuit board in a pattern shown on the photograph in figure 1.

The BC-630 silicone optical grease from Saint-Gobain (refractive index of 1.47) [20] is used to couple all optical components and to fill the volume inside the ceramic holders in front of the SiPMs. Note that the presence of the gap between the SiPMs and the lightguide, filled with the optical grease, increases the effective lightguide thickness to 2.1 mm.

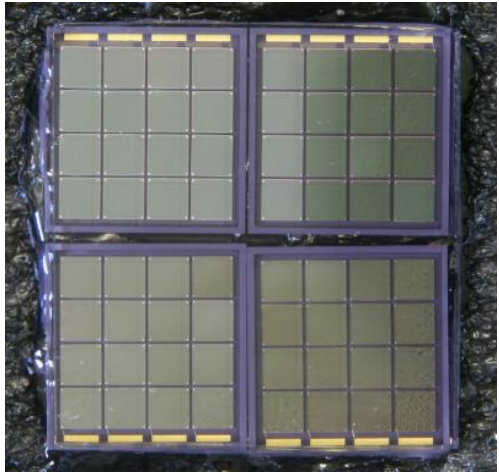

Figure 1. Photograph of the sensor array (four ArraySB-4 units from SensL) of the experimental prototype.

The readout system is based on the MAROC3 ASIC from Omega [21]. The chip has 64 low noise inputs with individually adjustable gains and two (slow and fast) signal processing circuits. The SiPM signals, shaped by the slow circuit and digitized with a 12 bit Wilkinson ADC, are transferred to a PC, where a pre-processing procedure is applied to subtract the pedestals recorded at a voltage slightly below the SiPM breakdown and to scale signals according to the channels-per-photoelectron values. Note that most of the events generated due to the natural radioactivity of LYSO result in saturation of the ADC and are filtered out.

The fast signal processing circuit of MAROC3 is used for event triggering. A trigger is generated if signal amplitude in any channel is above a given threshold, common for all 64 channels. To have approximately uniform triggering efficiency over the camera field of view, the preamplifier gains were adjusted using the following procedure: the camera was uniformly irradiated and only one channel at a time was allowed to trigger. The gain of this channel was adjusted to obtain a given triggering rate. The trigger level, common for all channels, was set high enough to avoid triggering on electronic noise but as low as possible to enable triggering for the events in the regions of the camera where the distance from the event to the center of the closest SiPM is larger than usual (see the cross-shaped area in figure 1). However, this compromise resulted in a somewhat lower triggering efficiency in the area of 1.5 mm by 1.5 mm in the center of the camera.

Experimental data were recorded using a $^{99m}$Tc source (gamma ray energy of 140 keV). Flood field irradiation data were collected by positioning the source at a distance of 1 m from the camera. A system of two lead collimators (apertures with the diameter of approximately 1.1 mm, spaced by 50 mm) was used to generate a pencil beam. The collimated source was installed on an XY positioning system (2.5 μm positioning precision) to record scans of the camera. Two masks were also used: a slit and a LIP logo (see section 3.3). Both masks were produced by milling 0.5 mm wide grooves in a 3 mm thick plate made of a bismuth-lead alloy. In the case of the LIP logo, to keep the mask elements together, the grooves are only 2.5 mm deep. In this study, the source and the collimators were always positioned on the PTFE reflector side of the camera.

## 2.3   Gamma camera simulations

Two models of compact gamma camera were simulated in this study. The first one is an "ideal" camera with a regular sensor array (figure 2, left). The camera has a 2 mm thick, 30 x 30 mm$^2$ LYSO(Ce) scintillator, which is coupled to a regular array of 8 x 8 SiPMs through a 2 mm thick lightguide of the same area as the scintillator. The SiPMs, each having an area of 3.16 x 3.16 mm$^2$, are positioned with a pitch of 3.8 mm.

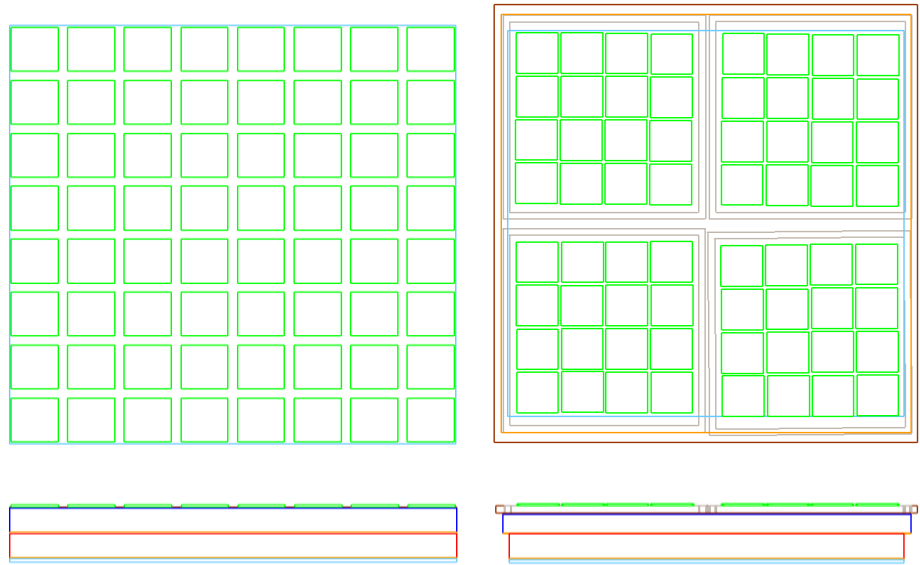

Figure 2. Two models of compact gamma camera simulated in this study. Left: a model of an "ideal" camera with a regular array of photosensors. Right: a camera configured to reproduce the design of the experimental prototype. The top and the bottom rows show the top and the side view, respectively. The SiPMs are green, the lightguides are blue, the scintillators are red, and the backplane reflectors are light-blue.

The scintillator's photon yield is set to 25 photons/keV for 140 keV gamma rays, the emission peak is assumed to be at 420 nm and the refractive index is set to 1.82 [18]. One plane of the scintillator is coupled to the PTFE reflector, while the other one is facing the SiPM array through the lightguide with the refractive index of 1.5, as for acrylic glass. The SiPMs characteristics are defined according to the MicroSB-30035 data sheet from SensL [19]: the photon detection efficiency of 31% at the emission peak of LYSO (420 nm), 4774 microcells and the dark count rate of $1 \times 10^7$ Hz.

The optical components of the camera are separated from each other by 0.1 mm thick layers of optical grease (refractive index of 1.47). To simulate the effect of black acrylic paint covering the edges in the prototype, the scintillator and the lightguide are positioned inside a medium with the refractive index equal to that of the paint (1.5). Because the refractive index of LYSO (1.82) is significantly larger than this value, reflection from the edges of the scintillator are strong (note the critical angle of 54 degrees). The photons exiting the scintillator and the lightguide on the lateral surfaces are considered absorbed. Thus, 100% absorption by the paint is assumed.

The second camera model is defined to more closely reproduce details of the design of the experimental prototype (figure 2, right). The main difference is the arrangement of the sensor array:

the SiPMs are grouped in four sub-arrays as in ArraySB-4 model from SensL: each sub-array hosts 4 by 4 SiPMs placed inside a ceramic holder. The individual SiPMs are positioned at the coordinates estimated from the photograph of the sensor array of the prototype (figure 1). Following the design of the ceramic holders, the active areas of the SiPMs are situated at a distance of 0.6 mm from the top plane of the lightguide. The gap between the lightguide and the SiPMs is filled with the optical grease. As in the prototype, the lightguide is 1.5 mm thick and has an area of 31 x 32.5 mm$^2$ due to the XY asymmetry of the ceramic holders. All other properties of the camera are the same as for the first model.

All simulations were performed using the ANTS2 package designed for simulation and experimental data processing for Anger camera type detectors. A detailed description of the package can be found in [22] and a brief one, summarizing the most important features relevant for simulation of the medical gamma cameras, in [16]. The package source code and the user manual are available online [23].

The simulations were performed to study detection of 140 keV gamma rays taking into account photoelectric effect and Compton scattering. The attenuation coefficients were taken from the XCOM database [24]. Electrons were assumed to deposit all their energy at the gamma ray interaction points. The intrinsic energy resolution of LYSO was assumed to be 13% [25] when generating scintillation photons. Optical photons were traced taking into account the refractive indexes of the media. The following *optical overrides* (see [22]) were defined: on crossing the interface from optical grease to the PTFE reflector, the photons had a fixed 95% chance to be scatted back following the Lambert's cosine law. On crossing the interface from optical grease to ceramic (SiPM holders) the probability of Lambertian reflection was set to 15%. The photons entering PTFE or ceramic were considered absorbed. Reflection on the interface from LYSO to optical grease (including full internal reflection) was simulated according to the Fresnel equations.

Gamma rays, generated in the direction normal to the scintillator were irradiating the camera from the PTFE reflector side. Event datasets were obtained by generating gamma rays uniformly over the scintillator area (flood field irradiation) or over Ø1 mm areas with 2.1 mm pitch to simulate detector scan with a pencil beam.

The SiPM signal formation was not simulated in detail: the output was given as the number of detected photons taking into account dark counts, which were generated assuming Poisson statistics with the average of one dark count per SiPM, corresponding to the experimental conditions.

## 2.4   Position reconstruction

Event position reconstruction was performed using the contracting grids method described in [16], implemented on a graphics processing unit (GPU). In short, for each event a regular grid of positions is defined, which is centered at the XY coordinates given by the centroid reconstruction. For each node of the grid, the algorithm evaluates the difference between the measured (or simulated) signals and the corresponding expected values given by the LRFs, assuming that the light was emitted from this node. The grid node, resulting in the best match between the measured (or simulated) signals and the expected from the LRFs, is selected and used then as the center of a finer grid covering the vicinity of this node. For the new grid the same number of nodes is used but the grid step is reduced by a given factor. The procedure is repeated until a sufficiently small (compared to the detector spatial resolution) grid step is reached. The estimate of the number of photons emitted during the event is calculated for each node as the ratio of two sums over all sensors: the sensor signals and the LRF values at that node [15].

Several modifications to the algorithm described in [16] have been made. Instead of the least squares approach applied before, the maximum likelihood algorithm assuming Poisson distribution of the number of photoelectrons was used to find the grid node which results in the best match. In contrast to the gamma camera model studied in [16], where it was possible to assume that the number of photoelectrons follows the normal distribution, for the compact gamma camera considered in this study, this assumption is not adequate due to smaller solid angles subtended by the active areas of the SiPMs. Furthermore, we have limited the number of SiPMs taken into account during reconstruction of an event: the sensors with centers situated farther than 10 mm from the position reconstructed by the centroid algorithm were ignored. This approach resulted in a more accurate reconstruction since the signals from those sensors were dominated by the SiPMs dark counts.

# 3   Results

## 3.1   Position reconstruction assuming axial symmetry of SiPM response

As mentioned in the introduction, the iterative response reconstruction technique was previously applied only for detectors with axially symmetric response of the photosensors. Therefore, the LRFs were always parameterized as functions of the distance from the sensor center. There are two reasons which make application of this parameterization questionable for the type of gamma camera considered in this study. First, it is the square shape of the sensors: one can expect that the profile of the LRF versus the distance from the sensor center strongly varies with the azimuthal direction. Second, it is the close proximity of the peripheral sensors to the crystal edge. In this region one can expect a significant contribution to the SiPM signals from the light reflected on the lateral surfaces of the scintillator (or the material covering these surfaces). This suggests that the axial symmetry is broken at least for the peripheral sensors.

An analysis of the spatial dependence of the solid angle subtended by a square-shaped area of size *S* for a point in a plane parallel to the area and situated at a distance *Z* from it shows that the presence of a lightguide with a sufficient thickness can mitigate the first problem. For example, for *Z* of *S*/3, *S*/2 and 2*S*/3, the maximum difference in the solid angle for different azimuthal directions but the same radial distance from the area center do not exceed 15%, 6% and 3%, respectively.

A Monte Carlo simulation was carried out in order to evaluate the effect of light reflection from the crystal edges. A $5\times10^5$ event dataset was obtained for the flood irradiation conditions of the camera model with the regular sensor array (figure 2, left). The event signals together with the known positions of the events were used to compute the spatial response of the sensors. The obtained responses for SiPMs situated at different distances from the crystal edge are shown in figure 3 as profiles versus the distance from the corresponding SiPM center. Every plot contains a set of 50 profiles for azimuthal directions regularly distributed over 2π. One can see that the responses of the peripheral SiPMs are indeed asymmetric. However, the SiPMs situated one row farther from the crystal edge already exhibit good axial symmetry, and the symmetry is practically perfect for the rest of the sensors.

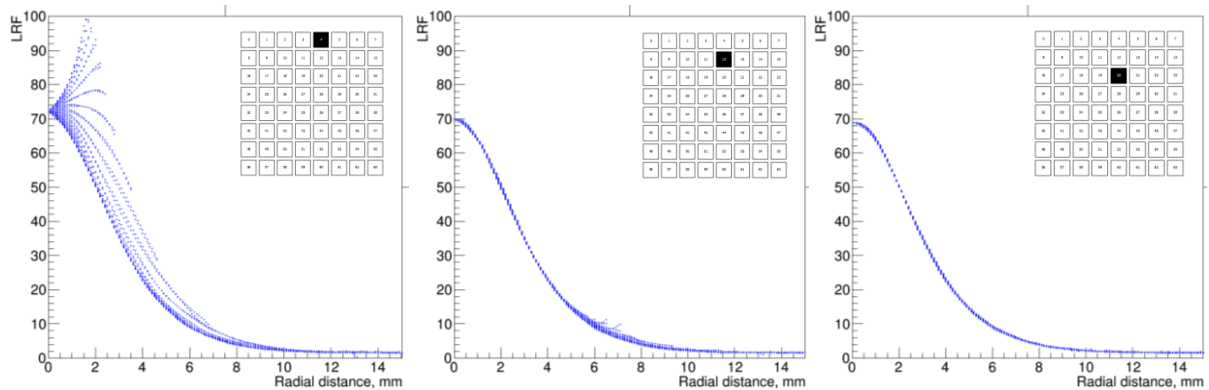

Figure 3. Profiles of the spatial response of the SiPMs as a function of radial distance from the SiPM center are shown for three SiPMs with different distances from the crystal edge. Each plot contains 50 profiles for azimuthal directions regularly distributed over 2π. The position of the corresponding SiPM is indicated in the top-right corner (black square). Note that the crystal area is 30 x 30 mm$^2$.

The results presented above indicate that parameterization of the sensor response using axially-symmetric LRFs may provide a good approximation for the most of the sensors except the ones closest to the scintillator edge. Using such parametrization, one can expect low level of distortions for the most of the field of view of the detector. However, in the regions close to the peripheral SiPMs distortions can be large.

To evaluate the distortion pattern for statistical reconstruction performed with axially-symmetric LRFs, the same flood field dataset was used to compute LRFs of the sensors using the axially symmetric parameterization based on B-splines [26] available in ANTS2. Since only the distance from the sensor axis to the event position is considered during the LRF computation, the resulting LRFs provide the expected sensor signal, averaged over all azimuthal directions, as a function of distance from the sensor center.

Since the positions of the simulated events are known exactly and the LRFs are computed directly, we refer to this type of LRFs as "direct" LRFs in order to distinguish them from iteratively reconstructed LRFs ("iterative" LRFs) introduced in the next section, which are obtained without knowledge of the true event positions.

The direct LRFs were used to reconstruct the positions of events obtained in a simulation of detector scan with a 1 mm pencil beam over a grid with 2.1 mm pitch. The resulting event density map and the corresponding true positions of the source (circles) are shown in figure 4. One can see that the distortions in the central region are negligible, and only at 12 mm from the scintillator center they become apparent.

The distortion pattern over the entire field of view of the camera can be better visualized using a distortion map (figure 5, left), which shows the difference between the true and the reconstructed event position in X direction, averaged over 0.5 x 0.5 mm$^2$ bins, plotted versus true event position. The distortion map was obtained by reconstructing event positions of another flood field dataset and using the same direct LRFs. Figure 5 also shows two profiles obtained from the distortion map.

One profile is along the diagonal direction from the bottom-left to the top-right corner, and the other one for the horizontal direction from left to right, passing the camera center. To reduce statistical fluctuations, the profiles show data averaged over 3 bins. The results demonstrate that the systematic distortions do not exceed 0.3 mm in the 29 x 29 mm² central area.

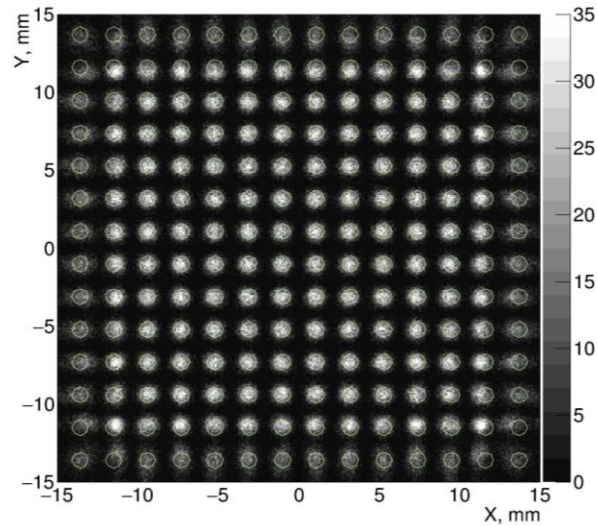

Figure 4. Demonstration that axial LRF parameterization is adequate for the compact gamma camera. LRFs were directly calculated from the simulated flood field dataset (no iterative procedure) using known true event positions. The density map of the reconstructed positions is superimposed with the circles indicating the true positions of the pencil beam source.

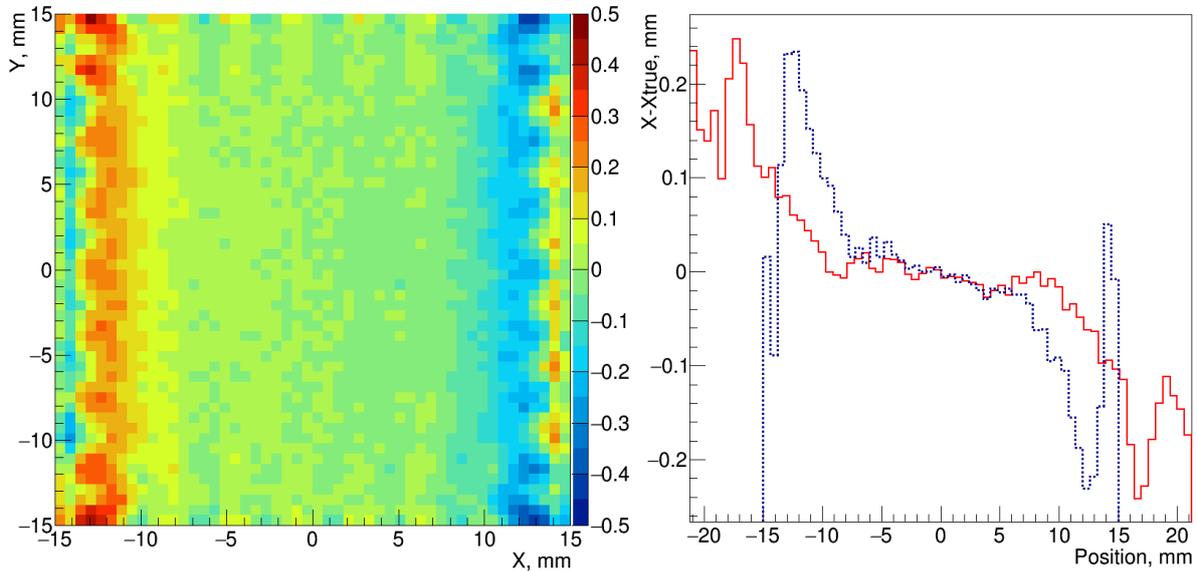

Figure 5. Distortion in the reconstructed positions, obtained using the direct (non-iterative) LRFs. Left: The distortion map, showing the difference between the true and the reconstructed X coordinates (color-coded, scale is in mm), averaged over a bin of 0.5 x 0.5 mm², and plotted versus true event position. Right: Two profiles, showing the data averaged over three bins for the horizontal (dotted, blue) and diagonal (red) directions (both through the camera center).

Figure 6 shows the average reconstructed event energy, as an XY map and, again, as the profiles in the diagonal and horizontal directions. The results demonstrate that the average reconstructed energy is flat in the central area, and does not fluctuate more that ±10% over more than 85% of the scintillator area.

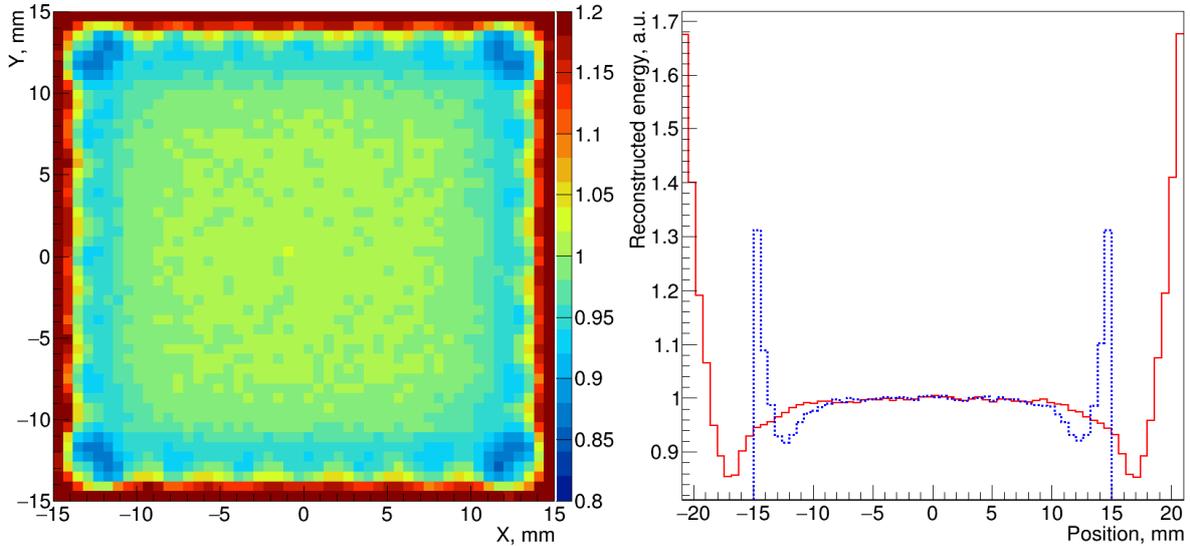

Figure 6. Average event energy reconstructed with the direct (non-iterative) LRFs versus true event position, shown as a color-coded map (left) and two profiles (right). Unitary energy corresponds to 140 keV. The profiles show the data averaged over three bins for the horizontal (dotted, blue) and diagonal (red) directions (both through the camera center).

## 3.2 Validation of the iterative method based on simulations

The iterative LRF reconstruction method was first applied for the "ideal" camera model (figure 2, left). Two types of datasets were obtained in simulations: one with flood field irradiation ($5\times10^5$ events) and the other with scans of the camera with Ø1 mm pencil beam (14 x 14 nodes with a pitch of 2.1 mm, 2500 events per node).

The LRF reconstruction[1] was performed using the flood simulation data. The initial guess on the LRFs was obtained from the centroid reconstruction. For the first three iterations a parameterization scheme with the LRF common for all sensors was used. The Gaussian blurring (see section 2.1) with sigma of 1 mm was applied after each iteration. After the third iteration the parameterization scheme was changed to the one in which all photosensors had individual LRFs and another eight iterations were performed. At this point a convergence was reached: the chi-square and the average deviation between the true and the reconstructed positions stopped to improve.

Figure 7 shows the images reconstructed with the centroid as well as with the maximum likelihood method using the LRFs obtained after one, four and eleven iterations. In the last image the true positions of the source are indicated with the circles. Note that this image is practically undistinguishable from the one obtained using the direct LRFs (see figure 4). A comparison of the distortion maps for the reconstructions with the direct and the iterative LRFs for a flood field dataset also shows a very similar level of distortions (figure 8). The map for the reconstruction with direct LRFs exhibit somewhat smaller distortions in the central area, while the map for the iterative LRFs shows slightly smaller distortions in the peripheral area.

---

[1] In contrast with section 3.1, where the known exact positions of the simulated events have been used to directly compute the LRFs, here this information is not used (as it would be for experimental data).

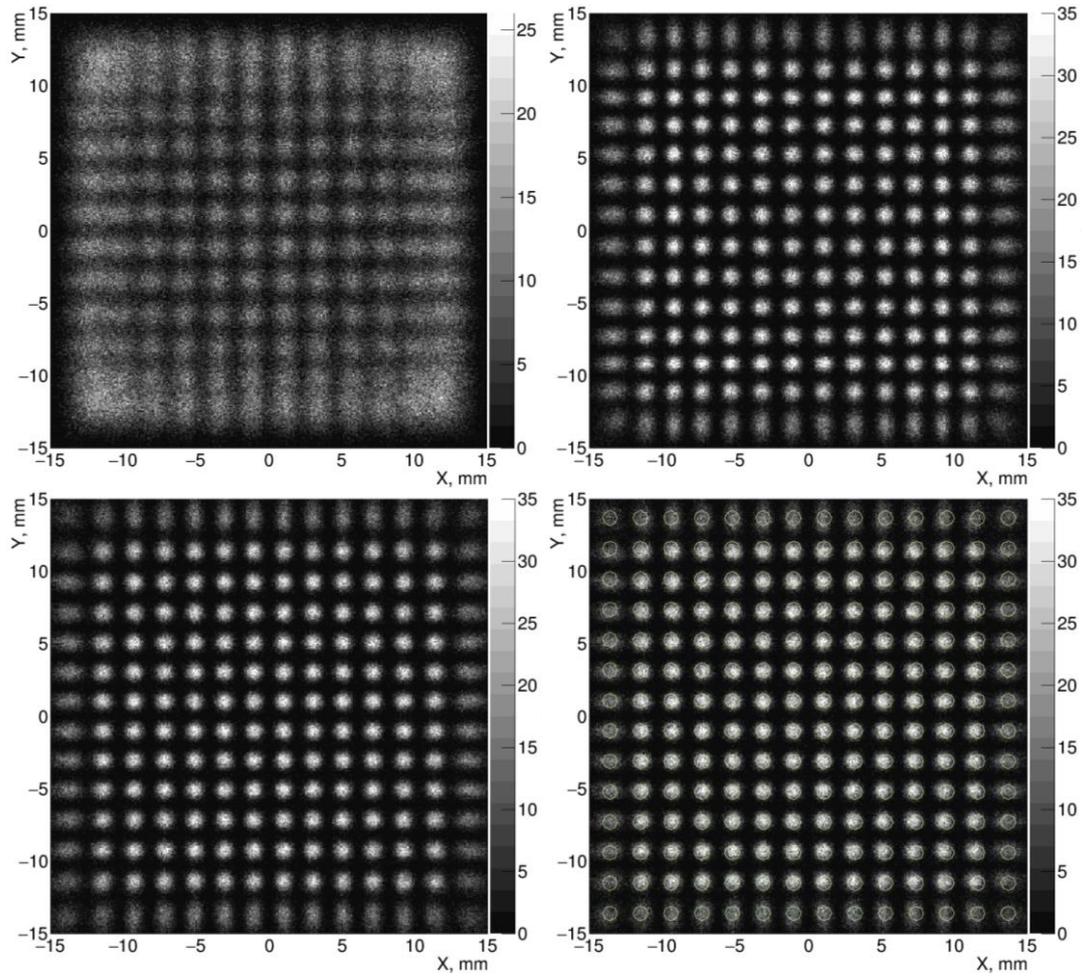

Figure 7. Density map of the reconstructed events for the scan simulation. The top-left image is obtained using the centroid reconstruction. All other images are reconstructed with the maximum likelihood method using the LRFs provided by the iterative method after one (top-right), four (bottom-left) and eleven iterations (bottom-right). In the last image the true positions of the source are indicated with the circles.

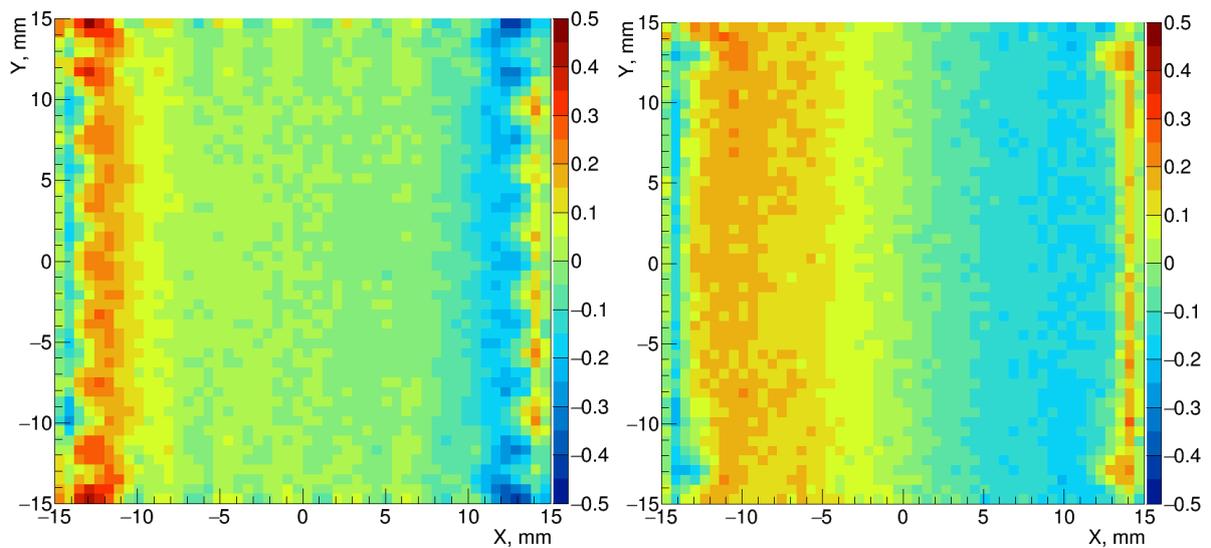

Figure 8. Distortion maps for the reconstruction performed with the direct and the iterative LRFs (left and right, respectively).

In the simulation described above all sensor had the same gains. However, this is not likely be the case when experimental data are used to reconstruct LRFs. Therefore, for practical applications, the method should be tolerant to a significant variation in the gains. The variations in the gains may appear, for example, due to continuous gain drift or errors introduced during the data pre-processing, when the signals are converted to the number of photoelectrons. To demonstrate the tolerance of the method, the sensor gains were scaled by random factors uniformly distributed from 0.5 to 1.5, and the simulations of the flood irradiation and the pencil beam scan were repeated.

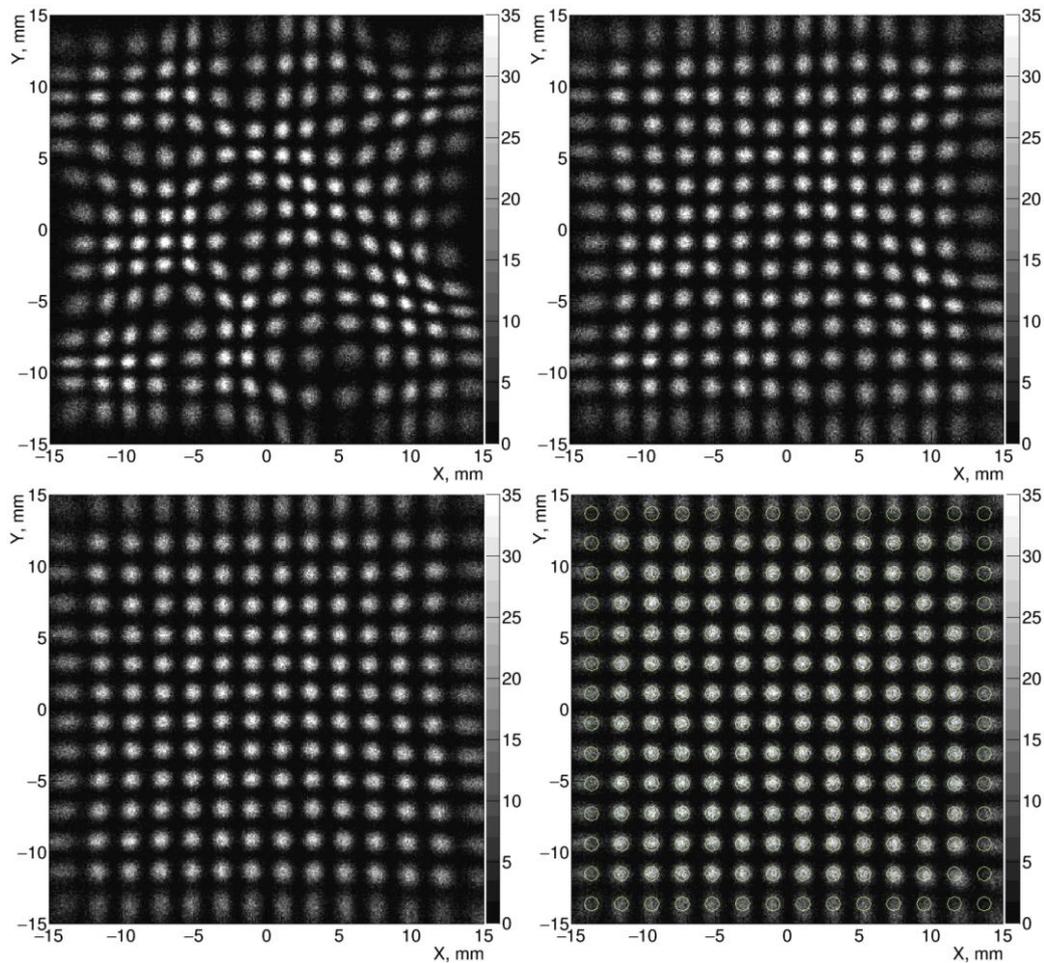

Figure 9. Density map of the reconstructed events for the scan simulation with random gains in the range from 0.5 to 1.5. The top-left image is obtained using the LRFs iteratively reconstructed for the simulation with unitary gains of all sensors. All other images are reconstructed with the maximum likelihood method using the LRFs obtained by the iterative method after one (top-right), four (bottom-left) and twenty iterations (bottom-right). In the last image the true positions of the source are indicated with the circles.

Instead of starting from the centroid reconstruction to provide the first guess on the LRFs, the final iterative LRFs obtained for the simulation with unitary gains of all sensors were used. Starting directly with the individual LRF parameterization scheme, 20 iterations were performed. A Gaussian blur with a sigma of 0.5 mm was applied after every iteration. Figure 9 demonstrates improvement of the reconstruction images during iterations. As one can see, although the initial guess results in a very distorted image, after 20 iterations the image quality is similar to the one obtained with the direct LRFs (figure 7). A slightly higher level of distortions in the last image of figure 9 is most likely explained by the fact that the maximum likelihood method assumed Poisson distribution of the sensor signals, while the dataset contained scaled data (number of photoelectrons multiplied with the relative sensor gain).

The next simulation was performed for the gamma camera representing the experimental prototype (figure 2, right). Using the same approach, it was shown that the iterative method can provide LRFs very similar to the direct ones starting from the initial guess based on the centroid reconstruction (see figure 10). Because of the larger inter-sensor distances in some areas of the detector, the level of distortion is somewhat larger (figure 10, bottom-right) than the one obtained for the camera with fully regular array, but the maximum deviation still does not exceed 0.4 mm.

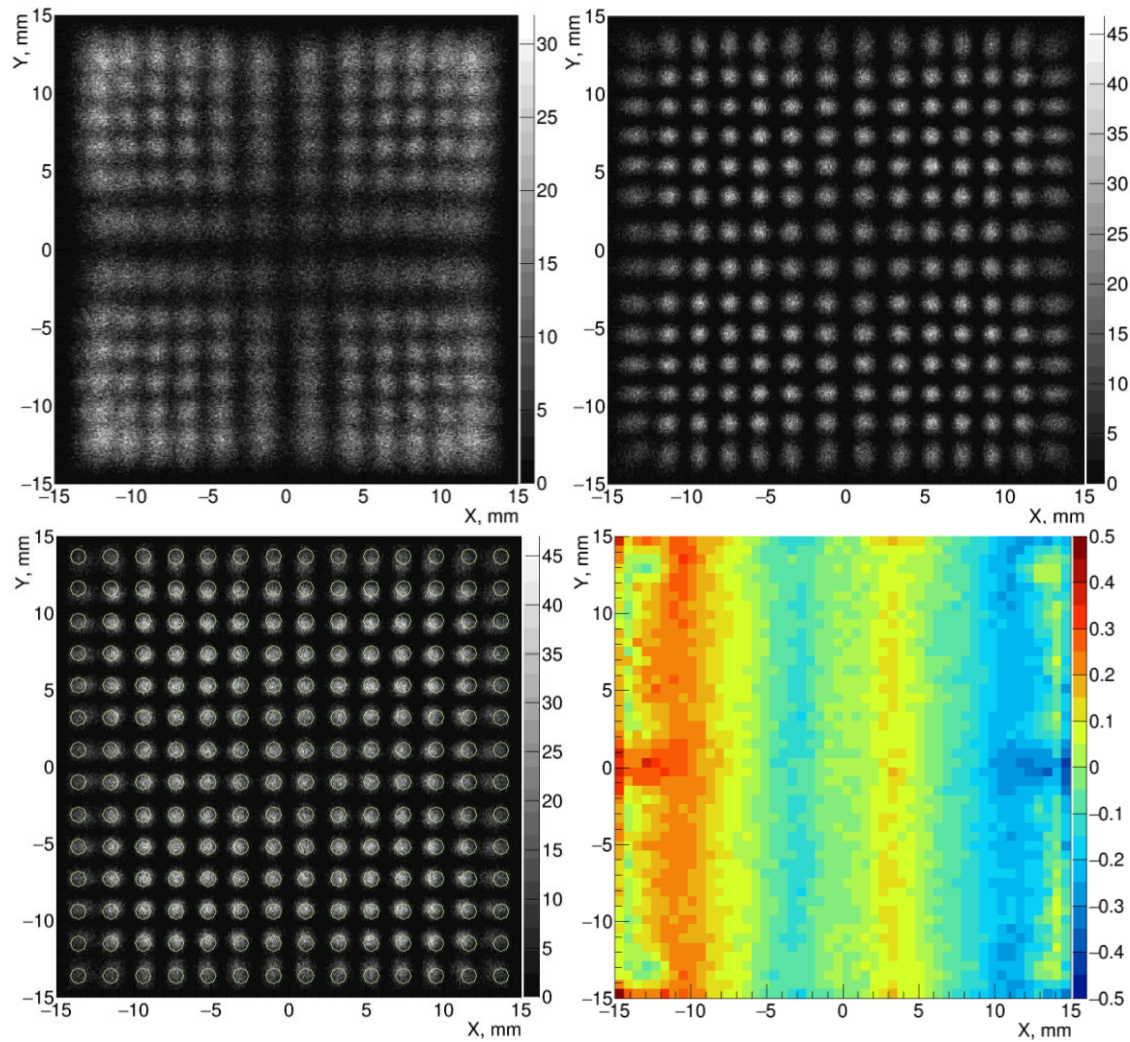

Figure 10. Results for the simulation of the camera model representing the experimental prototype. The images reconstructed with the centroid (top-left) and the maximum likelihood method using the LRFs provided by the iterative method after one (top-right) and fourteen iterations (bottom-left). The true positions of the source are indicated with the circles. Bottom-right: Distortion map calculated for a flood field dataset.

### 3.3 Experimental validation of the iterative method

For experimental validation of the iterative LRF reconstruction method, flood irradiation and camera scan datasets were recorded with the camera prototype. Flood irradiation data contained $5 \times 10^5$ events, and the camera scans were recorded with a ∅1.1 mm pencil beam over 14 x 14 nodes with a 2.1 mm pitch, acquiring 25000 events at each node.

The LRF reconstruction was performed using the flood field data with the initial guess on the LRFs obtained from the centroid reconstruction. Then, using common LRF parameterization, four iterations were performed, followed by eight more iterations with individual LRF parameterization, until the average chi-square value over the dataset stopped improving. After each reconstruction step a Gaussian blur with a sigma of 1 mm (common LRFs) or 0.5 mm (individual LRFs) was applied. During the first four iterations, the LRF calculation was performed ignoring events reconstructed outside the central area of 28 x 28.5 mm$^2$.

The scan datasets were used to evaluate the quality of the LRFs obtained in the iterative procedure. Figure 11 shows examples of the images obtained from the scan dataset with the centroid reconstruction as well as with the reconstruction using the maximum likelihood method with the LRFs calculated after one, three, five and twelve iterations. One can see a significant improvement of the image after just one iteration. The following iterations gradually improve the regularity of the inter-node distances (most noticeable in the central area). The circles superimposed on the last image indicate the true positions of the source. The match between the reconstructed and the true positions is similar to the one obtained in the simulations: over the 24 x 24 mm$^2$ area the difference between the true X (and Y) center position of the pencil beam source and the mean of the reconstructed event coordinate given by Gaussian fit is less than 0.3 mm. Note that the resolution for the experimental data is worse compared to the simulations due to the fact that the signal formation and electronic noise were not simulated. Also, the experimental data contain background events absent in the simulations.

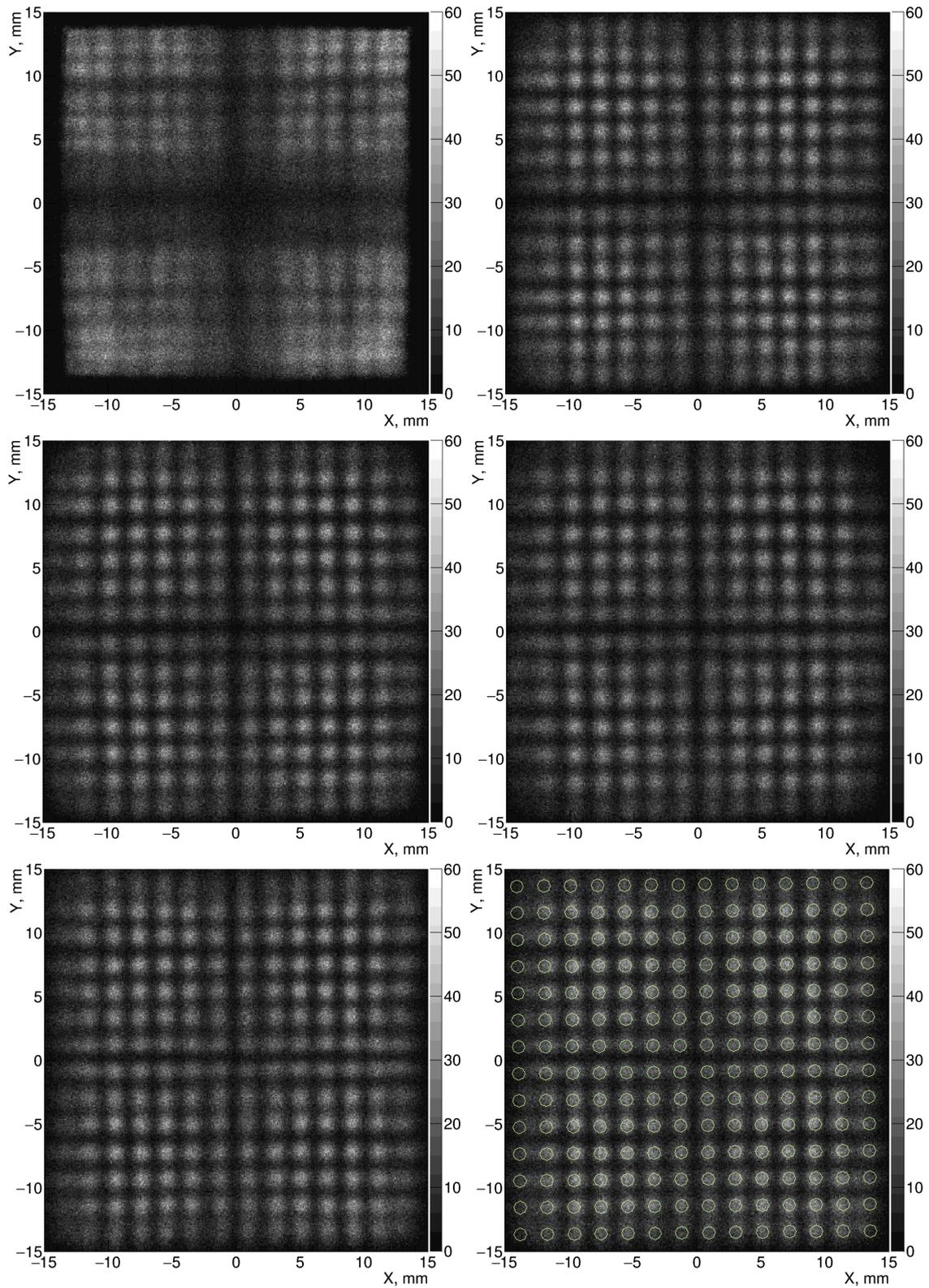

Figure 11. Iterative LRF reconstruction with experimental data. The centroid reconstruction (top-left) and the maximum likelihood reconstruction with the LRFs obtained after one (top-right), three (middle-left), five (middle-right) and twelve (bottom left and right) iterations. The circles in the last image indicate the true positions of the pencil beam source.

To give another visual demonstration of the low level of distortions provided by the reconstruction with the iterative LRFs, the datasets recorded with a LIP-logo mask (figure 12) and a 0.5 mm wide slit collimator (figure 13) were reconstructed.

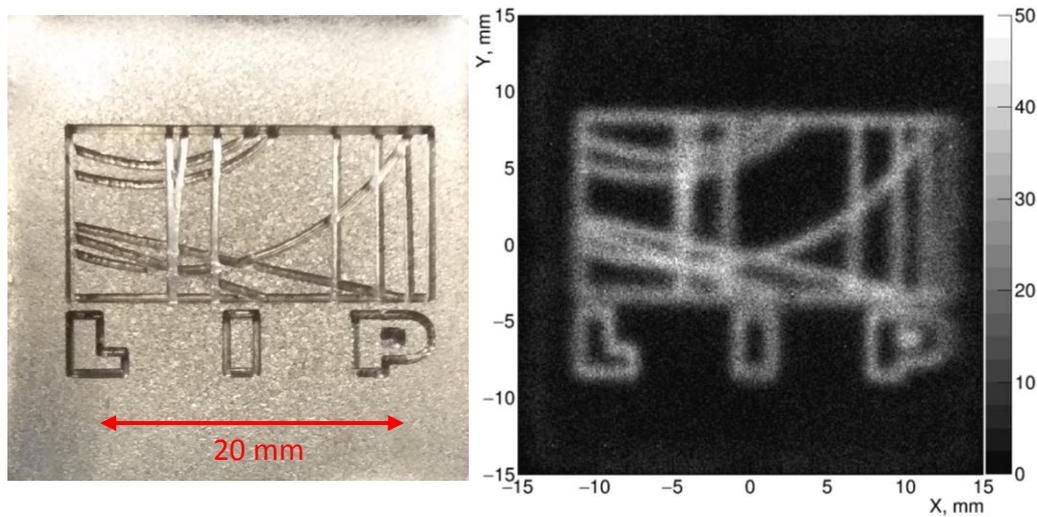

Figure 12. LIP-logo mask photograph (left) and the image reconstructed with the maximum likelihood algorithm using the iteratively reconstructed LRFs (right). The groove width is 0.5 mm.

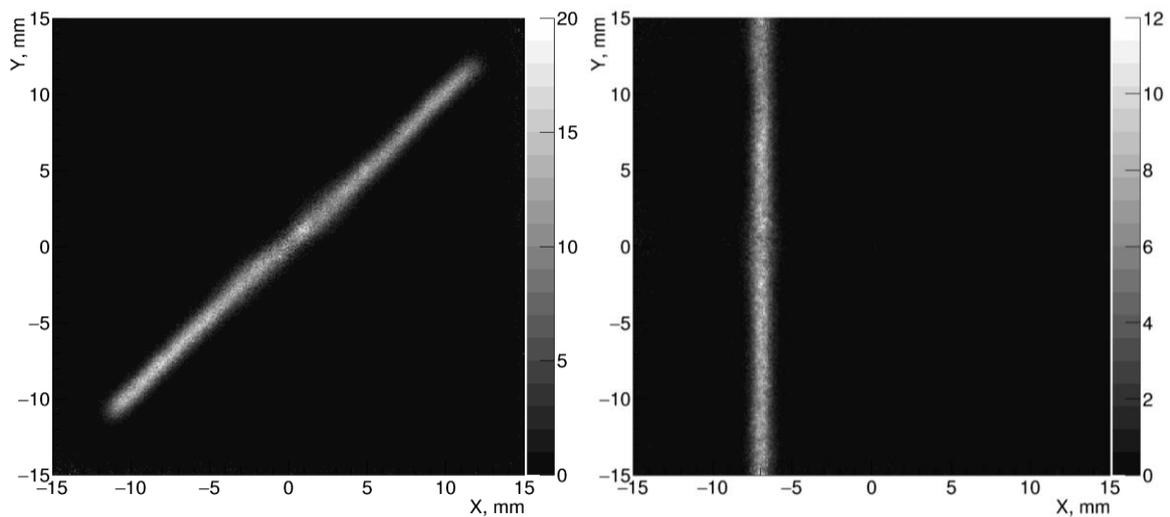

Figure 13. Images of the slit reconstructed with the maximum likelihood algorithm using the iteratively reconstructed LRFs. The slit width is 0.5 mm.

The image of the diagonal slit shows a drop in the triggering efficiency in the center of the camera (see section 2.2). The image of the vertical slit demonstrates worsening of the spatial resolution at the center due to longer inter-sensor distance in this area (see figure 1). The X-projection of the 0.5 mm vertical slit in the range of Y from -13 mm to -3 mm is shown in figure 14. The Gaussian fit results in a FWHM of 0.93 mm. If all available range in Y is used, the FWHM increases to 1.0 mm due to the contribution from the central area.

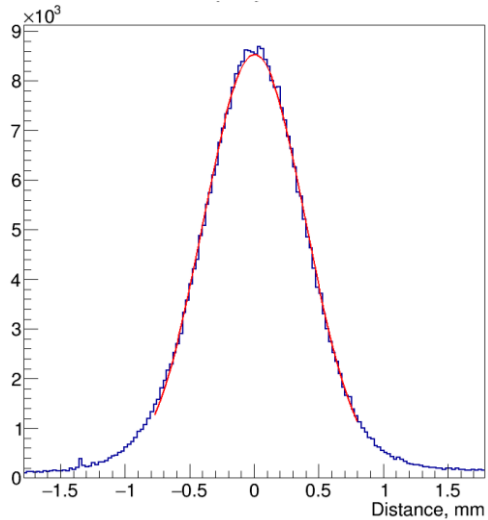

Figure 14. X-projection of the reconstructed event density for a dataset recorded with a 0.5 mm slit. A Gaussian fit results in a FWHM of 0.93 mm.

The average reconstructed energy map is shown in figure 15 (left). Similarly to the simulations (figure 6), the average reconstructed energy shows a sharp increase at the periphery of the camera. There is also a ~10% drop in the central "cross" area. The energy spectrum for the events reconstructed in the central area of 24 x 24 mm$^2$ is shown in figure 15 (right). A Gaussian fit of the peak results in energy resolution of 37%.

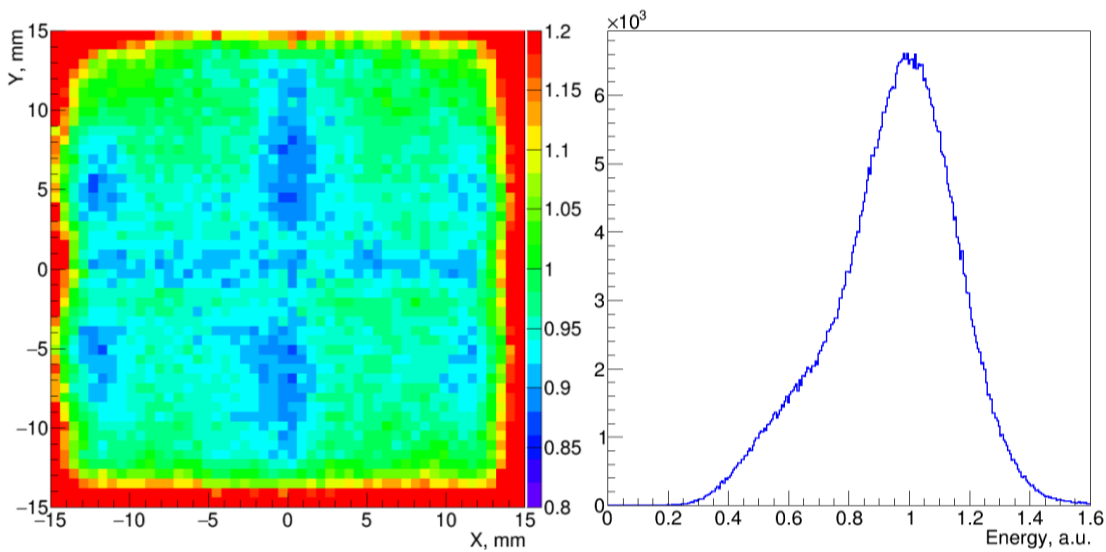

Figure 15. Map of the reconstructed energy, averaged over 0.5 x 0.5 mm$^2$ pixel area (left) and the reconstructed energy spectrum (right) for the flood irradiation data. The spectrum is shown for the central area of 24 x 24 mm$^2$. A contribution to the spectrum from the natural radioactivity of $^{176}$Lu of LYSO results in a peak at energy of 2.5 a.u. with a height less than 60 counts which extends to about 3.5 a.u. (not shown here).

## 4 Discussion and conclusions

The simulation results presented in section 3.1 demonstrate that for the compact gamma camera design analysed in this study it is possible to perform statistical reconstruction using axially symmetric LRFs for all photosensors including those situated close to the crystal edges. The maximum distortion in both X and Y directions in the images reconstructed with the maximum likelihood method using these LRFs does not exceed 0.3 mm over the whole crystal area.

This fact allowed us to apply the iterative LRF reconstruction method initially developed for Anger camera type detectors with axially symmetric response of the photosensors. The capability of this technique to evaluate LRFs from flood field calibration was demonstrated in simulations for two versions of the camera design, as well as for experimental data recorded with the camera prototype. The obtained LRFs were used in statistical reconstruction and the resulting images have shown a low level of distortions. For simulations, the difference between the reconstructed and the true positions in X and Y directions, averaged over 0.5 x 0.5 mm$^2$ bin area, does not exceed 0.2 mm in the central area of 22 x 22 mm$^2$ and 0.4 mm for the rest of the camera. For experimental data, the maximum difference in X (and Y) coordinate between the center of the pencil beam and the mean reconstructed event coordinate, given by Gaussian fit, is below 0.3 mm for the central area of 24 x 24 mm$^2$.

The procedure of iterative LRF reconstruction is rather quick. On a general purpose PC, equipped with a 3.4 GHz Intel Core i7 processor and an Nvidia GTX 970 GPU board, one iteration for a dataset of 5×10$^5$ events takes about five seconds (1.2 s for the reconstruction and event filtering, and the rest for the calculation of the LRFs). Therefore, even for 20 iterations, the total time required to reconstruct a set of 64 LRFs from a flood field calibration dataset is below 2 minutes.

The main drawback of the presented method is that the distortions, both in the reconstructed positions and energy, significantly increase in the peripheral region (2 mm from each side of the 30 mm square crystal). Replacement of the black paint used in this study to cover the edges of the scintillator with one having refractive index close to that of LYSO should partially mitigate this problem.

The energy resolution obtained for the camera prototype (37% FWHM) is worse than the resolution of 29% reported for a similar detector [9], utilizing digital SiPMs. We consider this to be mainly the consequence of a higher level of electronic noise in our readout system. Detector optimization, targeting spatial and energy resolution, will be addressed in a separate study.

# 5 Acknowledgements

This work was carried out with financial support from the Fundação para Ciência e Tecnologia (FCT) of Portugal through the project grants IF/00378/2013/CP1172/CT0001 and PTDC/BBB-BMD/2395/2012 (co-financed with FEDER), as well as from the Quadro de Referência Estratégica Nacional (QREN) in the framework of the project Rad4Life.